\documentclass[12pt,preprint]{aastex}

\slugcomment{}

\shorttitle{Sequestration of ethane in the cryovolcanic subsurface of Titan}

\shortauthors{Olivier Mousis \& Bernard Schmitt}

\begin{document}

\title{Sequestration of ethane in the cryovolcanic subsurface of Titan}

\author{
Olivier~Mousis\altaffilmark{1},
\& Bernard~Schmitt\altaffilmark{2}
}

\altaffiltext{1}{Universit{\'e} de Franche-Comt{\'e}, Institut UTINAM, CNRS/INSU, France; olivier.mousis@obs-besancon.fr}

\altaffiltext{2}{Universit{\'e} Joseph Fourier, Laboratoire de Plan{\'e}tologie de Grenoble, CNRS/INSU, France; Bernard.Schmitt@obs.ujf-grenoble.fr}

\begin{abstract}
{Saturn's largest satellite, Titan, has a thick atmosphere dominated by nitrogen and methane. The dense orange-brown smog hiding the satellite's surface is produced by photochemical reactions of methane, nitrogen and their dissociation products with solar ultraviolet, which lead primarily to the formation of ethane and heavier hydrocarbons. In the years prior to the exploration of Titan's surface by the {\it Cassini-Huygens} spacecraft, the production and condensation of ethane was expected to have formed a satellite-wide ocean one kilometer in depth, assuming that it was generated over the Solar system's lifetime. However, {\it Cassini-Huygens} observations failed to find any evidence of such an ocean. Here we describe the main cause of the ethane deficiency on Titan: cryovolcanic lavas regularly cover its surface, leading to the percolation of the liquid hydrocarbons through this porous material and its accumulation in subsurface layers built up during successive methane outgassing events. The liquid stored in the pores may, combined with the ice layers, form a stable ethane-rich clathrate reservoir, potentially isolated from the surface. Even with a low open porosity of 10\% for the subsurface layers, a cryovolcanic icy crust less than 2300 m thick is required to bury all the liquid hydrocarbons generated over the Solar system's lifetime.}
\end{abstract}


\section{Introduction}
\label{sec:intro} 
Any discussion on the reservoirs of ethane condensate at the surface of Titan should take into account the recently revised condensation rate of $5.9 \times10^{-14}$ g cm$^{-2}$ s$^{-1}$ for this molecule, consistent with Cassini CIRS observations (Atreya et al. 2006). This lower ethane condensation rate implies a reduction of the depth of the initially expected satellite-wide ocean to a value of $\sim$155 m, if methane was continuously released over Titan's life. This value is even smaller if we consider Titan's recent thermal evolution models supporting the idea that the methane outgassing has occurred episodically (Tobie et al. 2006). If the actual atmospheric methane is outgassing at the current rate since only $\sim$0.6 gigayears (Gyr) (Tobie et al. 2006), the depth of the hypothetical global ethane ocean reaches no more than $\sim$20 m. Indirect evidence has been obtained for the presence of lakes at high latitudes in the northern hemisphere during several Cassini Radar flybys of Titan (Stofan et al. 2007; Lorenz et al. 2008). But with a likely current inventory of $3 \times 10^4$--$3 \times 10^5$ km$^3$ of liquid (Lorenz et al. 2008), these lakes are not expected to store more than about 20\% of the produced ethane. This value can even be lower if we consider alternative photochemical models which lead to the condensation of twice as much ethane and predict the presence of substantial amounts of liquid propane (Vuitton \& Yelle 2005; Vuitton et al. 2008).

The apparent deficiency in liquid ethane on Titan has been interpreted as supporting the theory that ethane mostly condenses onto smog particles forming different types of thick deposits, including dunes and dark areas (Hunten 2006). However, the haze production rate is only 8 to 50\% of the ethane value (Atreya et al. 2006). With all the liquid and solid condensates taken into account, the liquid-to-solid volume ratio ranges between 2.5 and 23 (see Appendix \ref{liquid}). No granular material, even extremely microporous, can adsorb more than 30\% of its volume in its pores at saturation (see Appendix \ref{haze}). Any liquid above this value will wet the saturated haze, up to the point where the plastic and then the liquid limits of the material are reached. With liquid-to-solid volume ratios larger than 2.5, these limits are largely exceeded and the rheological properties of such wet material are incompatible with the stability of the observed dunes. In the most favorable case, the dunes and the wetter dark flat area could probably retain up to 20\% of the total liquid inventory, but the remaining 80\% should escape the haze deposits and migrate to other reservoirs (see Appendix \ref{haze}). 

An alternative interpretation is that a liquid ocean would fill the empty space of the upper crust consisting in a 10,000 m regolith layer generated by impacts during Titan's early history (Kossacki \& Lorenz 1996). After 4.5 Gyr of Titan's evolution, the porosity of this layer would decrease to 1--4\%, allowing the incorporation of a liquid ocean with an equivalent depth of $\sim$100--400 m (Kossacki \& Lorenz 1996). However, these calculations do not consider the porosity closing off and the depth below which it becomes isolated from the surface. On Earth, the porosity close-off in pure water ice is $\sim$10\% for Greenland's ice cap (Schwander et al. 1993), but for the finer grain size of the regolith this value is probably even higher. The resulting amount of trapped liquid is then lower than predicted. Moreover, the {\it Cassini} Radar and optical images show that the satellite surface is geologically young (Elachi et al. 2005). Therefore, it is unlikely that large extents of the early regolith remain currently in contact with the atmosphere since the deposition of photochemical debris, pluvial erosion or deposition, tectonic processes or cryovolcanism may have contributed to their burial.

None of the aforementioned ethane trapping scenarios is fully supported by the {\it Cassini-Huygens} observations of Titan (Atreya et al. 2006; Elachi et al. 2005; Stofan et al. 2007). However, the detection of cryovolcanic features (Elachi et al. 2005; Sotin et al. 2005; Lopes et al. 2007), together with the theoretical estimates of strong resurfacing rates (about 50 m of ``cryolava'' deposited per 10$^6$ yr; Elachi et al. 2005), lead us to propose a geological process that solves the ethane deficiency issue in a manner which is in agreement with our current knowledge of Titan: the incorporation of liquid hydrocarbons in the porous cryovolcanic subsurface. However, note that, at present, no extensive cryovolcanic deposits have yet been clearly identified on the surface of Titan but the presence of some distinct cryovolcanic features makes our scenario worth developing.

\section{Cryovolcanism on Titan}

We investigate the conditions of such incorporation by considering two different mechanisms of cryovolcanism and methane delivery:

\paragraph{Ascent of liquid from the subsurface ocean.} The methane coming from the saturated deep ocean (Tobie et al. 2006; Fortes et al. 2007), is transported close to the surface of Titan in ammonia-water pockets which erupt through the ice shell and lead to cryovolcanism (Mitri et al. 2006). The release of dissolved methane creates gas-rich eruptions and very porous cryovolcanic materials since, even with only 1\% of dissolved methane, the gas volume expelled largely exceeds that of the cryovolcanic ice\footnote{The methane equivalent gas/liquid volume ratio is $\sim$380 at 1.5 bar and 250 K (Lide 2002), namely a temperature value consistent with that within the icy crust.}.

\paragraph{Destabilization of clathrates in the ice shell of Titan.} In this case, clathrates of methane stored in the close subsurface are destabilized by ascents of hot thermal plumes and melt (Tobie et al. 2006). The cryolava expelled to the surface of Titan releases the large amount of methane initially trapped in clathrates. A very porous ice is produced from this clathrate decomposition (Schmitt 1986).\\

In both cases, a highly porous icy material is generated, probably similar to basaltic lava flows\footnote{ On Earth, the porosity of basaltic rocks is 10--50\% (Saar \& Manga 1999).}. Since the cooling of the cryolava is expected to take less than one year to decrease down to Titan's surface temperature (Lorenz 1996), it should be fast enough to allow the preservation of most of the porosity created by the methane release.

\section{Trapping of liquid hydrocarbons in the subsurface layers}

The liquid composition formed on Titan's surface can be inferred if it is considered in thermodynamic equilibrium with the atmosphere and that the organic materials mixed with the fluid have only minor effects on this equilibrium. With a methane mole fraction of $\sim$$4.9 \times 10^{-2}$ measured near the surface by {\it Huygens} (Niemann et al. 2005), the predicted mole fractions of methane, ethane and nitrogen are $\sim$0.35, $\sim$0.60 and $\sim$0.05 in the liquid, respectively (Dubouloz et al. 1989). We adopt these fractions as the nominal composition of the liquid currently present at and below the surface of Titan. 

We now consider two different Titan evolution scenarios and estimate the equivalent depth of a global liquid layer condensed on its surface: over a 0.6 Gyr period of cryovolcanism (Tobie et al. 2006), and over the lifetime of the Solar system (4.55 Gyr). In the first case, the thickness of the liquid layer is $\sim$30 m, including 21, 8 and 1 m of ethane, methane and nitrogen, respectively. In the second case, the thickness increases to 225 m, with 155, 62 and 8 m of ethane, methane and nitrogen, respectively. If the upper layers of Titan's subsurface are mostly constituted by an homogeneous deposit of cryolavas with a mean porosity of 25\%, icy crust thicknesses of only $\sim$120 and 900 m are required to bury these oceans for the two scenarios. Even if the porosity is 10\% (Schwander et al. 1993) (the close-off porosity is probably even lower due to the large grain size of the cryolava and the colder Titan temperatures), ice crust thicknesses of only 300 and 2250 m deep, respectively, are required.

The porosity is not expected to evolve significantly in $\sim$20 Myr within the top few kilometers of Titan's crust (Kossacki \& Lorenz 1996). Hence, since photolysis irreversibly converts Titan's atmospheric methane to heavier hydrocarbons in $\sim$10 Myr (Lunine et al. 1989), all ethane produced and condensed at the surface has sufficient time to percolate in Titan's porous cryovolcanic subsurface before the closure of its porosity. The state and stability of the column of liquid accumulated in the porous crust will then depend on the thermal profile with depth. In particular the conditions for clathrate formation and stability may be reached over some depth range.

Presuming that the liquid at a given depth is in thermal equilibrium with the ice surrounding the pores, we consider two different temperature profiles derived from Titan's interior models and covering its overall thermal history (Tobie et al. 2006). Assuming that the ice porosity remains open within the first 5 km of the icy shell, the two hydrostatic pressure-vs-temperature curves in a column of liquid are calculated\footnote{The pressure as a function of depth in a column of methane-ethane-nitrogen fluid that may exist in the subsurface of Titan can be calculated assuming a high open porosity generated by active cryovolcanism and a complete filling of the column up to the surface. The pressure produced in the pores at a given depth by a column of fluid is given by the hydrostatic equation $P(z) = P(z_0) + \int_0^Z \rho_{mix}~g~dz$, where $P(z_0)$ is the atmospheric pressure at the ground level, $\rho_{mix}$ is the liquid density ($\sim$516.3 kg m$^{-3}$) for the nominal mixture composition and $g$ is the gravitational acceleration ($\sim$1.35 m s$^{-2}$).} and compared to the stability curves of clathrates in Fig. 1. The equilibrium pressure curve of the multiple guest clathrate (hereafter MG clathrate) is determined for the nominal liquid composition. The equilibrium pressure curves of methane, ethane and nitrogen single guest clathrates are determined by fitting the available laboratory data and their equations are of the form log $P_{eq} = A/T + B$, where $P_{eq}$ and $T$ are the partial equilibrium pressure (bars) and the temperature (K) of the considered species, respectively. Table 1 shows the values of constants $A$ and $B$ from our fits to laboratory measurements. Equilibrium pressure curves for MG clathrate formed from the liquid mixture, can be expressed as (Hand et al. 2006): 

\begin{equation}
P_{eq,MG} = \left [\sum_i \frac{y_i}{P_{eq,i}} \right ]^{-1}
\end{equation}

\noindent where $y_i$ is the mole fraction of the component $i$ in the fluid phase. Figure 1 displays the dissociation pressures of methane, ethane, nitrogen and MG clathrates as a function of temperature. At a given temperature, these clathrates are stable at pressures equal or higher than their equilibrium pressures. They are also more stable than the liquid with corresponding composition provided sufficient water ice is available. In both cases, the column of liquid is located within the thermodynamic stability domains of ethane, methane and MG clathrates. The methane-ethane-nitrogen liquid filling the pores of Titan's icy crust thus likely forms, at all depths, a MG clathrate with the available water ice.

\section{Discussion}
Although the hydrostatic stability of the liquid column is achieved at any depth for both types of temperature profiles when the column is filled up to the surface (hydrostatic pressure always larger than the equilibrium vapour pressure of the liquid), this is not the case for a shallow column of liquid accumulated at great depths, and thus at higher temperatures. For example, if the porosity close-off currently occurs at 3 km depth the vapour pressure at the bottom of the liquid column (123 K) is larger than 2 bars and requires at least 300 m of liquid to stabilize. For porosities of 10--25\%, a minimum equivalent ocean depth of 30--75 m is necessary, larger than the total amount of liquid produced in the scenario of recent methane outgassing. In such cases the liquid may ascend the column under the vapor pressure and erupt. However the much colder upper crust layers will probably recondense the boosting gas before it reaches the surface. Clathrate formation provides a more efficient way to stabilize the stored methane-ethane-nitrogen mixture at great depth. Its formation also greatly reduces the equilibrium gas pressure and thus stabilizes the remaining liquid phase. With our current knowledge of clathrate formation, the amount of liquid trapped in the form of MG clathrate remains difficult to estimate. Indeed, the extent of clathrate formation from the liquid phase can be limited by the very slow (and poorly known) kinetics at these low temperatures and the availability of water ice to clathration around the pores. Finally, the formation process of clathrates strongly reduces the porosity of the crust. With a volume expansion of the clathrate structure of $\sim$20\% compared to that of water ice, its formation in an ice layer with an initial porosity of 25\% reduces this quantity to 10\%, namely the close-off value observed in terrestrial ice caps (Schwander et al. 1993). So clathrate formation may well induce its self isolation by closing the pore network that allowed the liquid hydrocarbons to percolate the ice down to these depths.

Long periods of active cryovolcanism on Titan, as previously suggested by Tobie et al. (2006), should lead to the continuous build up of consecutive layers of cryolava. Under the effect of increasing pressure and temperature, the porosity of the deepest and older layers should reach the close-off value, thus isolating the stored liquid or clathrate from the porous network connected to the surface. Hence, ethane and methane that are firmly trapped in the form of MG clathrates within the subsurface cannot escape by diffusion from the deepest layers, whatever their subsequent thermal history. Only cryovolcanic events, or cracks, may destabilize these stored materials.

Finally, given that the lakes of Titan and the haze produced in its atmosphere are expected to trap only a small fraction of the produced liquid hydrocarbons, and the absence of evidence for the presence of large areas of ancient regolith generated during the Late Heavy Bombardment epoch (Gomes et al. 2005), this leads us to believe that the sequestration of ethane in porous cryolava is the main mechanism still acting on Titan at a global scale. This statement does not impede that the other trapping mechanisms may compete locally on some areas of Titan's surface. Thus, the dunes of Titan can indeed retain a small fraction of the condensed hydrocarbons. The presence of wet surfaces (Zarnecki et al. 2005) on Titan and the possibility that lakes communicate with a subsurface liquid table (Stofan et al. 2007) may be also the result of their contact with a saturated porous cryolava layer.

\acknowledgements
We thank G. Tobie and V. Vuitton for information on their models. P. Duval and G. Moreels are also acknowledged for their helpful comments. This work was supported by the International Space Science Institute (ISSI) and the French Centre National d'Etudes Spatiales (CNES).

\appendix

\section{Liquid-to-solid ratio of the condensable species}
\label{liquid}
The condensation rate of the haze has been estimated from models and observations to range between 0.5 and $3 \times 10^{-14}$ g cm$^{-2}$ s$^{-1}$ (McKay et al. 2001; Wilson \& Atreya 2003). On the other side, the condensation rate of liquid ethane is derived to be $5.9 \times 10^{-14}$ g cm$^{-2}$ s$^{-1}$ (Atreya et al. 2006), leading to an ethane to haze mass ratio ranging from 2 to 12. With liquid propane on one side and solid acetylene and all other solid condensates (Wilson \& Atreya 2004) added on the other side, the lower limit of this ratio is reduced to 1.3. Assuming conservatively a bulk density of the individual aerosol particles similar to that of polyacetylene (0.63 g cm$^{-3}$) and considering that ethane accounts to 60\% of the whole liquid phase (CH$_4$ + C$_2$H$_6$ + C$_3$H$_8$ + N$_2$) (see text) then the bulk volume ratio of the liquid to the solid phases (haze + solid condensates) deposited at the surface ranges between 2.5 and 23. A recent chemical model (Vuitton \& Yelle 2005; Vuitton et al. 2008) gives a typical volume ratio (C$_2$H$_6$ + C$_3$H$_8$ liquid condensates)/(haze + C$_2$H$_2$ + all solid condensates) of 3.3 that leads to a total liquid-to-solid ratio of 5.5 (including methane). This value is well within the previous range, although with significantly larger ethane and propane production rates.

\section{adsorption and rheological properties of ``wet'' haze}
\label{haze}
For spherical haze monomers with 0.05 $\micron$ radius, their specific surface area (Surface/Volume ratio) is 60 m$^2$ cm$^{-3}$ and typical adsorbant-to-adsorber volume ratio (including condensation in pores of all sizes formed when these particles aggregate) is 0.1--0.3 near saturation ($P$ = 0.95 $P_s$, the saturation pressure of the liquid). Even in extreme cases of highly porous solids (with ``equivalent'' specific surface area in the range 300--500 m$^2$ cm$^{-3}$), this ratio rarely reaches a value over 0.5 (see e.g. Gregg \& Sing 1982). This case may possibly occur for microporous monomer haze particles. But laboratory simulations produce spherical tholins particules without such a porous structure (Szopa et al. 2006). In addition it is also incompatible with the scattering properties of the particules assumed for the DISR data inversion, supposed to be spherical and compact (Tomasko et al. 2005). Any liquid above this 0.1--0.3 saturation value will begin to ``wet'' the saturated haze, up to the point where the liquid phase dominates the physical and rheological properties of the ``soil''. 

The transition between a cohesive particulate material and a plastic one (that can flow) depends on the material and on the liquid content. The amount of liquid at this transition is called the ``Plastic or Atterberg limit''. On Earth, this transition occurs for most natural materials for volumetric water contents (water/material bulk volume ratio) ranging from a few percents up to 50\% in the extreme cases of some clays (including interlayer water). The liquid limit is also almost always below 80\%. For pure organic materials these values are generally low ($<$ 20\%) and the ``liquid limit'' not far above the plastic limit (small ``plastic index''). Aggregates of compact sub-micron sized particules of organic ÒtholinÓ materials can adsorb and condense in its porous structure ~10--20\% of its bulk volume in liquid and can probably store no more than another 10--20\% of liquid in the macropores (between the aggregate particles) before reaching the plastic limit. With a liquid-to-solid volume ratio in the 2.5--23 range, the undifferentiated material is certainly well above the liquid limit. Such high to very high values should produce materials ranging from some kind of fluid mud to a liquid with only a few percents of aerosol particles in suspension. Thus the dunes, and their slopes, are not consistent with the rheological behaviour of such materials. The dark flat area on Titan, such as at the {\it Huygens} landing site, are consistent with wet surfaces, but not liquid. So they may retain locally up to 20--50\% of liquid. Averaged over Titan's surface both dunes and flat haze deposits could probably retain no more than 20\% of the total liquid inventory.

\clearpage
\begin{figure}
\resizebox{\hsize}{!}{\includegraphics[angle=0]{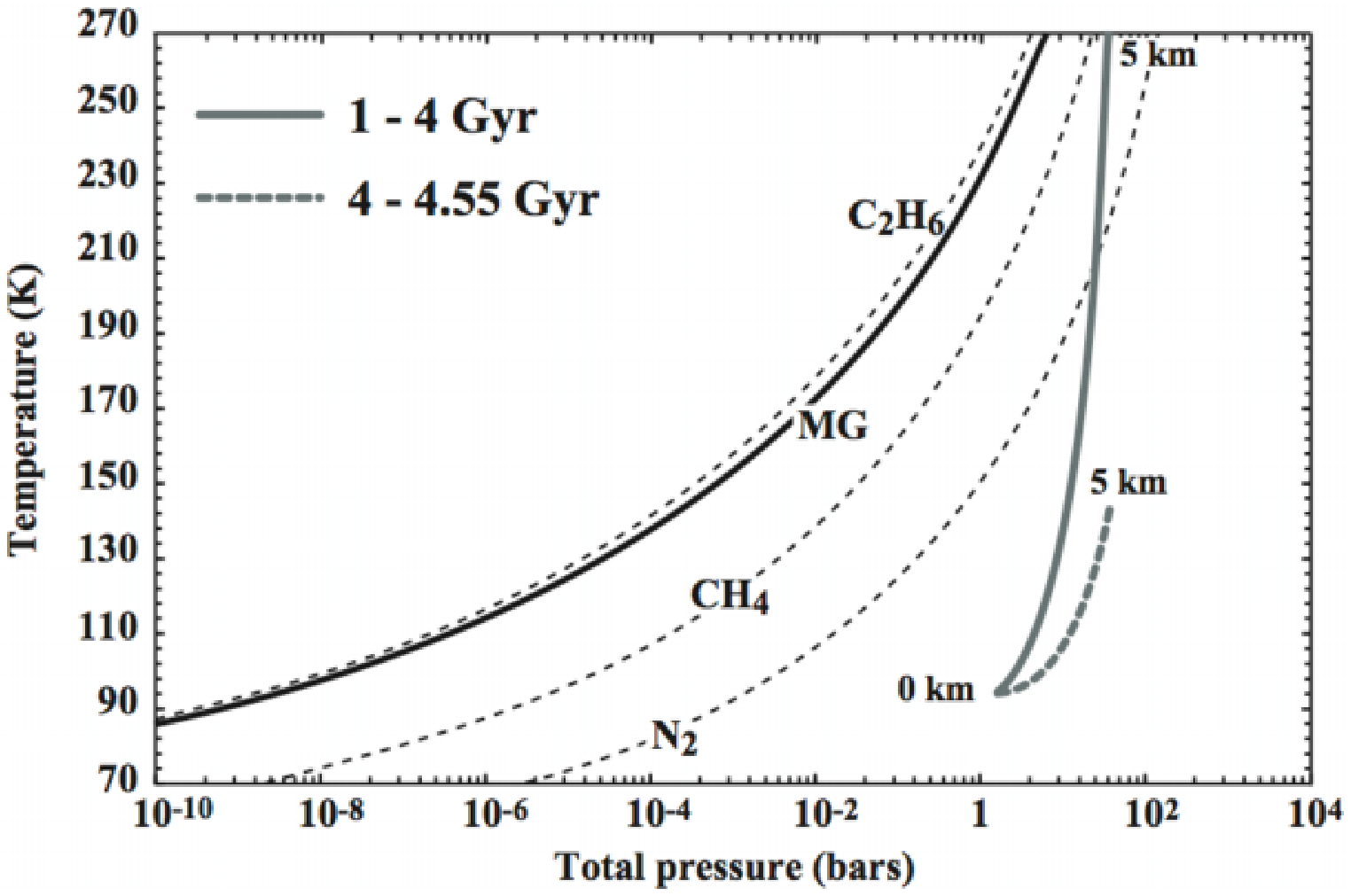}} 
\caption{From left to right: equilibrium pressures of C$_2$H$_6$, multiple guest (MG in the figure), CH$_4$ and N$_2$ clathrates as a function of temperature (black curves). The hydrostatic pressure within a column of CH$_4$-C$_2$H$_6$-N$_2$ liquid is also represented as a function of local temperature in the porous subsurface of Titan at two different periods of its thermal history. The first temperature profile (grey curve) corresponds to the 1--4 Gyr period of Titan's evolution. During this period, the upper water ice crust is conductive and its thickness does not exceed $\sim$5 km. A linear temperature profile can then be constructed between an assumed temperature of $\sim$94 K at the surface and $\sim$270 K at the inner ice-ocean interface. The second temperature profile (dotted grey curve), which is valid at epochs later than 4 Gyr, postulates the existence of a $\sim$50 km thick icy crust. In the outer conductive layer of the icy crust, about 15 km deep, the temperature varies linearly between $\sim$94 K and $\sim$250 K. In our calculation, we only consider the first 5 km of the icy crust since it can largely contain all the liquid hydrocarbons generated over Titan's life (see text).} 
\label{clat}
\end{figure}

\clearpage
\begin{table}[h]
\caption[]{Parameters of the equilibrium curves of the considered single guest clathrates. }
\begin{center}
\begin{tabular}{lccc}
\hline
\hline
\noalign{\smallskip}
Molecule 		&  	$A$  		& 	$B$  		& Reference \\	
\hline
\noalign{\smallskip}
Ethane		& 	-1366.70	& 	5.70		& Sloan (1998) \\
Methane		&  	-951.23	&  	4.89		& Lunine \& Stevenson (1985) \\
Nitrogen   		&    	-728.58 	&      	4.86		& Lunine \& Stevenson (1985) \\
\hline
\end{tabular}
\tablecomments{$A$ is in K and $B$ is dimensionless.}
\end{center}
\label{fit}
\end{table}

\end{document}